\documentclass[twocolumn,preprintnumbers,nofootinbib]{revtex4}
\usepackage{graphicx}
\usepackage{amssymb}
\usepackage{color}
\usepackage{enumerate}
\def\beq{\begin{equation}}
\def\eeq{\end{equation}}
\def\beqn{\begin{eqnarray}}
\def\eeqn{\end{eqnarray}}

\renewcommand{\texttt}{{}}
\newcommand{\be}{\begin{eqnarray}}
\newcommand{\ee}{\end{eqnarray}}

\begin{document}

\title{Super-renormalizable 
Higher-Derivative 
Quantum Gravity} 

\author{Leonardo Modesto}
\affiliation{Perimeter Institute for Theoretical Physics, 31 Caroline St., Waterloo, ON N2L 2Y5, Canada}

\date{\small\today}

\begin{abstract} \noindent
In this paper we study perturbatively an 
extension of the Stelle higher derivative 
gravity involving an infinite number of derivative terms. 
We know that the usual quadratic action is renormalizable 
but 
is not unitary because of the presence of a ghost in the theory 
(pole with negative residue in the propagator). 
The new theory is instead 
ghost-free since an entire function (or form factor) is introduced in the model without involving 
new poles in the propagator. 
The local high derivative theory 
is recovered expanding the entire functions to the lowest order in the mass scale of the theory. 
Any truncation of the entire function gives rise to 
unitarity violation. 
The theory is divergent at one loop and finite from two loops upwards: 
the theory is then super-renormalizable.
Using the modified graviton propagator, we demonstrate the regularity of the gravitational potential in $r=0$.

\end{abstract}

\maketitle


One of the biggest problems in theoretical physics is to find a theory that is able to merge 
general relativity and quantum mechanics. 
Some recent papers \cite{BM, modesto} 
introduced 
a different action principle for gravity 
to make up for the shortcomings of Einstein theory.
The theory fulfills 
a synthesis of minimal requirements: 
(i) classical solutions must be singularity-free;  
(ii) Einstein-Hilbert action should be a good approximation of the theory at a much smaller energy scale than the Planck mass;  
(iii) the spacetime dimension has to decrease with the energy in order to have 
a complete quantum gravity theory in the ultraviolet regime; 
(iv) the theory has to be perturbatively renormalizable at quantum level 
(this hypothesis is strongly related to the previous one), 
(v) the theory has to be unitary, with no other 
pole in the propagator in addition to the graviton, 
(vi) spacetime is a single continuum of space and time and in particular the 
Lorentz invariance is not broken. 
This approach to quantum gravity is partly inspired by the Cornish and Moffat papers \cite{Moffat}.

The theory we are going to introduce 
is well defined perturbatively at the quantum level.
Similarly, at the classical level the gravitational potential \cite{BM}, 
black hole solutions \cite{modesto, ModestoMoffatNico, NS} and the cosmological solutions are singularity free \cite{BM}. 
The Lagrangian is a ``nonlocal" extension of the renormalizable quadratic Stelle theory \cite{Stelle}
and it has the following general structure,  
\be
\! \mathcal{L}_g = -\sqrt{|g|} 
\left[\frac{2 R}{ \kappa^{2}} + R_{\mu \nu}  \gamma_1(\Box_{\Lambda})  R^{\mu \nu} + R \,\gamma_2( \Box_{\Lambda})  R \right] \! ,
\label{theory}
\ee
where 
the two ``form factors" $\gamma_{1,2}(\Box_{\Lambda})$ are ``entire functions" of the covariant D'Alembertian operator, $\Box_{\Lambda} = \Box/\Lambda^2$, $\Lambda$ is an invariant mass scale
and $\kappa^2 = 32 \pi G_N$.
The non locality only involve positive powers of the D'Alembertian operator since the two form 
factors are entire functions. 
The theory is not unique, but all the freedom present in the action can be read in 
the two functions $\gamma_{1,2}$ \cite{efimov, Krasnikov, Tombo}. 
Such functions must be interpreted in analogy 
with the interaction of a photon with a nucleon. 
Therefore, the form factors for gravity $\gamma_{1}( \Box_{\Lambda})$ and $\gamma_{2}( \Box_{\Lambda})$
can be measured experimentally.


We define the form factors $\gamma_{1,2}(\Box)$ 
in terms of two entire functions $h_2$ and $h_0$ 
which will appear in the spin two and spin zero part of the propagator, respectively 
\be
&& 
\gamma_1(\Box_{\Lambda}) \equiv  
 h_2( - \Box_{\Lambda} ) - \beta_2 \,  ,
\\
&& \gamma_2(\Box_{\Lambda}) \equiv 
   \frac{\beta_2}{3} + \beta_0   -   h_0 ( - \Box_{\Lambda}) 
- \frac{h_2( - \Box_{\Lambda}  ) }{3}   \, . \nonumber 
\ee
The entire functions $h_2$ and $h_0$ both depend on 
 the covariant 
D'Alembertian operator $- \Box_{\Lambda} \equiv - \Box/\Lambda^2$. 
 The complete Lagrangian is 
\be
\mathcal{L} = \mathcal{L}_g +  \mathcal{L}_{\rm GF} + \mathcal{L}_{\rm GH}, 
\ee
where the gauge fixing and ghost Lagrangian terms are
\be
 \mathcal{L}_{\rm GF} + \mathcal{L}_{\rm GH}= -\frac{1}{2 \xi} F^{\mu}  \omega( - \Box_{\Lambda}^{\eta})  F_{\mu} + 
\bar{C}^{\mu} M_{\mu \nu} C^{\nu} \label{gaugec}.
\ee
The operator $\Box^{\eta}_{\Lambda}$ encapsulates the D'Alembertian of the flat fixed background $\eta_{\mu\nu}$,
whereas  
$F_{\mu}$ is the gauge fixing function with the weight functional $\omega$.
$\bar{C}^{\mu}, C^{\mu}$ are the ghosts 
fields and 
$M^{\tau}_{\alpha} = F^{\tau}_{\mu \nu} D^{\mu \nu}_{\alpha}$. 
The gauge fixing function $F^{\tau}_{\mu \nu}$ and the operator $D^{\mu \nu}_{\alpha}$ will be defined 
shortly.
%
To avoid ghosts (states with negative norm) in the spectrum,
the two functions $h_2$ and $h_0$ must not be 
polynomial but ``entire", i.e. without poles or essential singularities.


%
%
%
%

We calculate now the graviton propagator. 
For this purpose 
we start by considering the quadratic expansion of the Lagrangian (\ref{theory}) in the graviton field
fluctuation without specifying the explicit form of the functionals $h_2$ and $h_0$. 
Following the Stelle paper \cite{Stelle}, we expand 
around the Minkowski background $\eta^{\mu \nu}$ 
in power of the graviton field $h^{\mu \nu}$  defined in the following way  
\begin{eqnarray}
\sqrt{ - g} g^{\mu \nu} = \eta^{\mu \nu} + \kappa h^{\mu \nu}. 
\label{graviton}
\end{eqnarray}
The form of the propagator depends not only on the gauge choice but also on the definition of the 
gravitational fluctuation \cite{Shapirobook}. 
The gauge choice is the familiar ``harmonic gauge"
$\partial_{\nu} h^{\mu \nu} =0$. 
In (\ref{gaugec}), $F^{\tau} = F^{\tau}_{\mu \nu} h^{\mu \nu}$ with $F^{\tau}_{\mu \nu} := \delta^{\tau}_{\mu}\partial_{\nu}$ and 
$D^{\mu \nu}_{\alpha}$ is the operator which generates the gauge transformations in the graviton 
fluctuation $h^{\mu \nu}$. Given the infinitesimal coordinates transformation 
$x^{\mu \prime} = x^{\mu} + \kappa \xi^{\mu},$ 
the graviton field transforms as follows 
\be
&& \delta h^{\mu \nu} = D^{\mu \nu}_{\alpha} \xi^{\alpha} 
= \partial^{\mu} \xi^{\nu} + \partial^{\nu} \xi^{\mu} - \eta^{\mu \nu}  \partial_{\alpha} \xi^{\alpha} \nonumber \\
&& \hspace{-0.5cm}
+ \kappa ( \partial_{\alpha} \xi^{\mu} h^{\alpha \nu} + \partial_{\alpha} \xi^{\nu} h^{\alpha \mu}
- \xi^{\alpha} \partial_{\alpha} h^{\mu \nu} - \partial_{\alpha} \xi^{\alpha} h^{\mu \nu} ).
\ee



Now we Taylor-expand the gravitational part of the action (\ref{theory}) to the second order 
in the gravitational perturbation $h^{\mu \nu}(x)$. 
In the momentum space the action, which is purely quadratic in the gravitational field, 
reads  
%
\begin{eqnarray}
\mathcal{L}^{(2)} = \frac{1}{4} 
h^{\mu \nu} (-k) 
 K_{\mu \nu \rho \sigma} \, h^{\rho \sigma}(k) + \mathcal{L}_{\rm GF} \, , 
\label{quadratic}
 \end{eqnarray}
where $\mathcal{L}_{\rm GF}$ is the gauge fixing Lagrangian at the second order in the graviton field 
\be
&&  \hspace{-1.1cm}
\mathcal{L}_{\rm GF} = \frac{1}{4 \xi}
 h^{\mu \nu} (-k) 
\Big( 
\,  \omega(k^2/\Lambda^2)  k^2 P^{(1)}_{\mu \nu \rho \sigma}(k) \nonumber \\
&& + 2  \omega(k^2/\Lambda^2) k^2 \} P^{(0 - \omega)}_{\mu \nu \rho \sigma}(k) \Big)
 h^{\rho \sigma}(k).
\label{GF}
\ee
The kinetic operator $K_{\mu \nu \rho \sigma}$ is defined by 
\begin{eqnarray}
&& \hspace{-0.4cm} 
K_{\mu \nu \rho \sigma} := 
  -  \bar{h}_{2}(z)    \, k^2 \, P^{(2)}_{\mu \nu \rho \sigma}(k)
%
+  \frac{3}{2} \, k^2 \,   \bar{h}_0(z)  \, 
P^{(0 - \omega)}_{\mu \nu \rho \sigma}(k) \nonumber \\
&& \hspace{-0.5cm}
+ \frac{k^2}{2} \,  \bar{h}_0(z) \, 
\{   P^{(0 - s)}_{\mu \nu \rho \sigma}(k) + \sqrt{3} [ P^{(0 - \omega s)}_{\mu \nu \rho \sigma}(k)
+ P^{(0 - s \omega)}_{\mu \nu \rho \sigma}(k) ] \}  , \nonumber 
 %
 %
\end{eqnarray}
where the quantities are thus explained 
\be
&& \bar{h}_2(z) := \beta - \beta_2 \kappa^2 \Lambda^2 z + \kappa^2 \Lambda^2 z \, h_2(z) , \nonumber \\
&& \bar{h}_0(z) := \beta - 6 \beta_0 \kappa^2 \Lambda^2 z + 6 \kappa^2 \Lambda^2 z \, h_0(z). 
\label{hbar}
\ee
While in (\ref{hbar}) $z := - \Box_{\Lambda}$, in (\ref{quadratic}) $z$ has to be identified 
with the D'Alembertian operator in flat spacetime $ - \Box^{\eta}_{\Lambda}$.
In the kinetic operator, we have introduced the 
projectors $P^{(2)}, P^{(1)},  P^{(0-s)}, P^{(0-s\omega)}, P^{(0-\omega s)} $ \cite{VN}. 
Using the orthogonality properties of the projectors 
we can now invert the kinetic matrix in 
(\ref{quadratic}) and obtain the 
graviton propagator. In the following expression, the graviton propagator is expressed 
in the momentum space according to the quadratic Lagrangian (\ref{quadratic}), 
\be
D_{\mu\nu\rho\sigma}(k) = D^{\xi=0}_{\mu\nu\rho\sigma}(k) + D^{\xi}_{\mu\nu\rho\sigma}(k),
\label{fullprop}
\ee
where the propagator in the gauge $\xi = 0$ and the pure gauge part of the propagator are  
\be
&& \hspace{-0.5cm}D^{\xi=0}(k) = \frac{- i}{(2 \pi)^4} \frac{2}{k^2 + i \epsilon} 
\Bigg( \frac{P^{(2)}}{ \bar{h}_2 \left( \frac{k^2}{\Lambda^2} \right) }
- \frac{2 P^{(0 - s)} }{ \bar{h}_0 \left( \frac{k^2}{\Lambda^2} \right)  } \Bigg) 
\nonumber \\
&& \hspace{-0.5cm}D^{\xi}(k) = 
\frac{ i}{(2 \pi)^4} \frac{ \, \xi }{(k^2 + i \epsilon) \, \omega \left(\frac{k^2}{\Lambda^2}   \right) } 
\Big( 2  P^{(1)} + 
 3 P^{(0 - s)} \nonumber \\
&& \hspace{1.1cm}
- \sqrt{3} \left( P^{(0 - s \omega)}  + 
P^{(0 - \omega s )}   \right) +  P^{(0 -  \omega)} \! \Big)
\label{propaImp}
\ee
and where we omitted the tensor indexes. 

%
The propagator in (\ref{propaImp}) explicitly depends on the two functions 
$\bar{h}_2$ and $\bar{h}_0$ that we are now going to explain 
in order to obtain the renormalizability of the theory along with unitarity.


We require the following general properties for the transcendental entire functions $h_i$ ($i = 2,0$)
\cite{Tombo}:
\begin{enumerate}
\renewcommand{\theenumi}{(\roman{enumi})}
\renewcommand{\labelenumi}{\theenumi}
\item $\bar{h}_i(z)$ is real and positive on the real axis and it has no zeroes on the 
whole complex plane $|z| < + \infty$. This requirement implies that there are no 
gauge-invariant poles other than the transverse massless physical graviton pole.
\item $|h_i(z)|$ has the same asymptotic behavior along the real axis at $\pm \infty$.
\item There exists $\Theta>0$ such that 
$$\lim_{|z|\rightarrow + \infty} |h_i(z)| \rightarrow | z |^{\gamma} \,\, , \,\,\,\,  
\gamma\geqslant 2$$ 
for the argument of $z$ in the cones 
\be
&& \hspace{-0.2cm} 
C = \{ z \, | \,\, - \Theta < {\rm arg} z < + \Theta \, , \,\,  \pi - \Theta < {\rm arg} z < \pi + \Theta \} , \nonumber \\
&&  \hspace{-0.2cm} 
{\rm for } \,\,\, 0< \Theta < \pi/2. \nonumber 
\ee
This condition is necessary in order to achieve the (supe-)renormalizability of the theory. The necessary 
asymptotic behavior is imposed not only on the real axis, 
but also in conic regions surrounding 
the real axis. In an Euclidean spacetime, the condition (ii) is not strictly necessary if (iii) applies.
\end{enumerate}
Given the above properties, let us study the ultraviolet behavior of the quantum theory.
From the property (iii) in the high energy regime, the propagator in momentum space goes as 
$1/k^{2 \gamma +4}$ (as it is evident when combining (\ref{propaImp}) with the property (iii)).
Accordingly, the $n$-graviton interaction has the 
same scaling, since it can be written in the following schematic way,
\be
&& \hspace{-0.25cm}
{\mathcal L}^{(n)} \sim  h^n \, \Box_{\eta} h \,\,  h_i( - \Box_{\Lambda}) \,\, \Box_{\eta} h \nonumber \\
&& \hspace{0.5cm}
\rightarrow h^n \, \Box_{\eta} h 
\,  ( \Box_{\eta} + h^m \, \partial h \partial )^{\gamma} \, 
\Box_{\eta} h , 
\label{intera}
\ee
where $h$ is the graviton field and $h_i$ is the entire function defined by the properties (i)-(iii). 
From (\ref{intera}), the superficial degree of divergence is 
\be
\hspace{-0.3cm} 
\delta = 4 L - (2 \gamma + 4) I + ( 2 \gamma + 4) V 
= 4 - 2 \gamma (L - 1).
\label{diver}
\ee
In (\ref{diver}) we used again the topological relation between vertexes $V$, internal lines $I$ and 
number of loops $L$: $I = V + L -1$. 
Therefore, 
only 
1-loop divergences survive and the theory is super-renormalizable\footnote{
A ``local" super-renormalizable quantum gravity with a large 
number of metric derivatives was for the first time introduced in \cite{shapiro}.}.
In this theory,  
 the quantities 
$\beta$, $\beta_2$, $\beta_0$ and eventually the cosmological constant $\lambda$ are renormalized. 
The Lagrangian reads: $\mathcal{L}_{\rm Ren} + \mathcal{L}_{\rm NL}$, where $\mathcal{L}_{\rm NL}$
is the 
Lagrangian containing the two entire functions $h_2, h_0$ and 
\be
\mathcal{L}_{\rm Ren} = - \sqrt{-g} 
\Big\{  \frac{\beta Z }{\kappa^2} R  + \lambda Z_{\lambda} 
- \frac{\beta_2 Z_2}{2}C_{\mu \nu \rho \sigma}^2 
+ \beta_0  Z_0 R^2 \Big\} . \nonumber
\ee 
$C_{\mu\nu\rho\sigma}$ is the Weyl tensor and $C^2 = 2(R_{\mu \nu} R^{\mu \nu} -  R^2/3 )$.
All the couplings must be understood as renormalized at an energy scale $\mu$. On the other hand,  
the functions $h_i$ are not renormalized. To better understand this point we can write 
the generic entire functions as series $h_i(z) = \sum_{r=0}^{+\infty} a_r z^r$.
For $r \geqslant 1$ there are no counterterms that renormalize $a_r$ 
because of the superficial degree of divergence (\ref{diver}). 
Only the coefficient $a_0$ can be renormalized; however, 
$\beta_2$ and $\beta_0$
already incorporate 
such renormalization. 
Therefore, the non-trivial dependence of the entire functions $h_i$ on their argument is preserved at quantum level.

By imposing the conditions (i)-(iii), we have the freedom to choose the following form for the entire functions 
$h_i$,
\be
&& h_2(z) = \frac{\alpha (e^{H(z)} -1) + \alpha_2 z}{\kappa^2 \Lambda^2 z}, \nonumber \\
&& h_0(z) = \frac{\alpha (e^{H(z)} -1)  + \alpha_0 z}{6 \kappa^2 \Lambda^2 z},
\label{hz}
\ee
for three general parameters $\alpha$, $\alpha_2$ and $\alpha_0$. 
$H(z)$ 
is an entire function that exhibits logarithmic asymptotic behavior in the conical region $C$.
Since $H(z)$ is an entire function, 
$\exp H(z)$ 
has no zeros in all complex plane for $|z|< + \infty$, according to the property (iii). 
If Taylor-expanded, $H(z)$  erases the denominator in $h_i(z)$
at any energy scale
and no inverse powers of the d'Alembertian operator appear in the action.

The entire function $H(z)$, which is compatible with the property (iii),
can be defined as 
\be
H(z) = \int_0^{p_{\gamma +1 }(z)} \frac{1 - \zeta(\omega)}{\omega} {\rm d} \omega \, , 
\label{Hz}
\ee
where the following requirements have to be satisfied:
\begin{enumerate}
\renewcommand{\theenumi}{\alph{enumi}.}
\renewcommand{\labelenumi}{\theenumi}
\item $p_{\gamma +1 }(z)$ is a real polynomial of degree $\gamma+1$ with $p_{\gamma +1}(0) = 0$,
\item $\zeta(z)$ is an entire and real function on the real axis with $\zeta(0) = 1$,
\item $|\zeta(z)| \rightarrow 0$ for $|z| \rightarrow \infty$ in the conical region $C$ defined in (iii). 
\end{enumerate}

Let us now peruse the unitarity of the theory. 
We assume that the theory is renormalized at some scale $\mu_0$. 
To avoid gauge-invariant poles 
in the bare propagator other than  
the physical 
graviton one, we have to set  
\be
\hspace{-0.4cm}
\alpha = \beta(\mu_0) \, , \hspace{0.3cm} 
\frac{\alpha_2}{\kappa^2 \Lambda^2} =  \beta_2(\mu_0) \, , \hspace{0.3cm}
\frac{\alpha_0}{6 \kappa^2 \Lambda^2} = \beta_0(\mu_0) .
\label{betaalpha}
\ee
%
If the energy scale $\mu_0$ is taken as the renormalization point, we get  
$\bar{h}_2 = \bar{h}_0 = \beta(\mu_0) \, \exp H(z):=\bar{h}(z)$. 
In the gauge $\xi = 0$, the propagator in (\ref{fullprop}) 
simplifies to 
\be 
&& \hspace{-1cm} 
D^{\xi = 0}(k) = 
%
\frac{- i}{(2 \pi)^4} \frac{e^{- H(k^2/\Lambda^2)}}{\alpha \, ( k^2 + i \epsilon) } 
\Big(  2 P^{(2)}
- 4 P^{(0-s)}\Big) .
\label{propalpha}
\ee
If we choose another renormalization scale $\mu$, the bare propagator acquires poles
that are canceled out in the dressed propagator. 
This result ensues from the renormalization group invariance.
When $h_2(z)=h_0(z)=0$ \cite{Stelle} the theory fails to be unitary and 
unitarity can be achieved only if $\beta_2=\beta_0=0$, which corresponds 
to the Einstein non renormalizable action.


An explicit example of entire function 
$H(z)$ 
compatible with the properties (i)-(iii) can be easily constructed. 
There are of course many ways to choose $\zeta(z)$, but we focus here on the obvious exponential choice  
$\zeta(z) = \exp(- z^2)$, which satisfies property c. in a conical region $C$ with 
$\Theta =\pi/4$. 
The entire function $H(z)$ is the result of the integral defined in (\ref{Hz}), 
\be
&& 
H(z) = \frac{1}{2} \left[ \gamma_E + 
\Gamma \left(0, p_{\gamma+1}^{2}(z) \right) \right] + \log [ p_{\gamma+1}(z) ] \, , 
\nonumber \\
&& {\rm Re}( p_{\gamma+1}^{2}(z) ) > 0, 
\label{H}
\ee
where $\gamma_E \approx 0.577216$ is the Euler's constant and  
$\Gamma(a,z) = \int_z^{+ \infty} t^{a -1} e^{-t} \rm{d} t$ is the incomplete gamma function.  
Another equivalent expression for the entire function $H(z)$ is given by the following series
\be
\hspace{-0.5cm} 
H(z) = \sum_{n =1}^{+ \infty} ( -1 )^{n-1} \, \frac{p_{\gamma +1}(z)^{2 n}}{2n \, n!} \, , \hspace{0.2cm}
{\rm Re}( p_{\gamma+1}^{2}(z) ) > 0.
\label{HS}
\ee
If we choose $p_{\gamma+1}(z) = z^{\gamma +1}$, $H(z)$ simplifies to:
\be
&& 
H(z) = \frac{1}{2} \left[ \gamma_E + \Gamma \left(0, z^{2 \gamma +2} \right) \right] + \log (z^{\gamma +1}) \, ,
\nonumber \\
&& {\rm Re}(z^{2 \gamma +2}) > 0. 
\label{H0}
\ee
For $p_{\gamma +1}(z) = z^{\gamma +1}$ the $\Theta$ angle, which defines the cone $C$ of (iii),  
is $\Theta = \pi/(4 \gamma+4)$. 
%

We now wish to include a more general class of theories following Efimov's study on non-local interactions \cite{efimov}.
Let us consider the propagator in the following general form 
\be
D(z)^{\xi =0} = \frac{V(z)}{ z \, \Lambda^2}
\label{propgeneral}
\ee
(the notation is rather compatible with the graviton propagator 
and $z := - \Box_{\Lambda}$).

As shown by Efimov \cite{efimov}
, the nonlocal field theory is ``unitary" and ``microcausal"
provided that the following properties are satisfied by $V(z)$, 
\begin{enumerate}
\renewcommand{\theenumi}{\Roman{enumi}}
\item
$V(z)$ is an entire analytic function in the complex
$z$-plane and has a finite order of growth $1/2 \leqslant \rho < + \infty$ i.e. $\exists \, b>0,c>0$ so that
\be
|V(z)| \leqslant c \, e^{b \, |z|^{\rho}}.
\ee
\item When ${\rm Re}(z) \rightarrow + \infty$ ($k^2 \rightarrow + \infty$), $V(z)$ 
decreases with sufficient rapidity. We can encounter the following cases:
\begin{enumerate}
\renewcommand{\labelenumii}{\alph{enumii}.}
    \item $V(z) = O\left(\frac{1}{|z|^a}  \right)$ ($a>1$) , 
    \item $\lim_{ {\rm Re}(z) \rightarrow +\infty} |z|^N |V(z)| =0$, $\forall \, N>0$. 
\end{enumerate}
\item  $[V(z)]^{*}= V(z^*)$.
\item $V (0) = 1$.
\item The function $V(z)$ can be non-negative on 
the real axis, i.e. $V(x) \geqslant 0$, $x = {\rm Re}(z)$. 
\end{enumerate}
Here 
we study the 
II.b. example of form factor, 
\be
V(z) = e^{- z^n} \,\, {\rm for} \,\,  n \in \mathbb{N}_+ \, , \,\, 
 \,\,  \rho = n < +\infty. 
\ee
When omitting the tensorial structure, the high energy propagator in the momentum space reads 
\be
 D(k) = e^{- (k^2/\Lambda^2)^n}/k^{2} \, .
 \label{expp}
\ee
The $n$-graviton interaction has the 
same scaling in the momentum space, 
as we have similarly 
highlighted before, 
\be
{\mathcal L}^{(n)} \sim 
h^n \, \Box_{\eta} h 
\,  \frac{\exp { \left(- \frac{\Box_{\eta}}{\Lambda^2} \right)^n } }{\Box_{\eta}} \, 
\Box_{\eta} h + \dots  \, , 
\label{intera2}
\ee
where ``$\dots$" indicates other sub-leading interaction terms 
coming from the
covariant D'Alembertian. 
Setting an upper bound to the 
$L$-loops amplitude, we find 
{\small
\begin{eqnarray}
&& \hspace{-0.4cm} \mathcal{A}^{(L)} \leqslant \int (d^4 k)^L \, \left(\frac{e^{-k^{2n}/\Lambda^{2n}}}{k^2} \right)^I \, 
\left(e^{k^{2n}/\Lambda^{2n}} k^2 \right)^V  \\
&& \hspace{-0.5cm} 
= \int (d k)^{4 L} \left(\frac{e^{-k^{2n}/\Lambda^{2n}}}{k^2} \right)^{I - V} 
\!\!\!\!\! = \int (dk)^{4 L} \left(\frac{e^{-k^{2n}/\Lambda^{2n}}}{k^2} \right)^{L-1} \!\!\!\!\! ,
\nonumber 
\end{eqnarray}
\hspace{-0.1cm}}
where in the last step we used again the topological identity $I = V+L-1$.
The $L$-loops amplitude is UV finite for $L>1$ and it diverges like ``$k^4$" for $L=1$.
Only 1-loop divergences survive in this theory.
%
%
%
The theory is then super-renormalizable, as well as unitary and microcausal 
\cite{efimov, Krasnikov}. 


%
Let us conclude by considering the modifications to the gravitational potential due to the form factor 
$V(z)$. 
Here we consider a static point particle source of energy tensor $T^{\mu}_{\nu} = {\rm diag}(- \rho,0,0,0)$ 
and $\rho = M \delta(\vec{x})$. 
Given the modified propagator $D(k) = 
V(k^2/\Lambda^2)/k^2$ (\ref{propgeneral}), 
the gravitational potential reads 
\be
&& \hspace{-0.5cm}\Phi(x) = - \frac{\kappa^2}{8} 
\int d^4 x^\prime \int \frac{d^4 k}{(2 \pi)^4} e^{ i k (x - x^\prime)} 
 \frac{V(k^2/\Lambda^2)}{ k^2} M \delta(\vec{x}^{\prime}) 
\nonumber \\
&& \hspace{0.35cm} = - \frac{\kappa^2 M }{8}  \int \frac{d^{3} k}{(2 \pi)^{3}} e^{ - i \vec{k} \cdot \vec{x}}
\frac{V(\vec{k}^2/\Lambda^2)}{ \vec{k}^2}. 
\label{int}
\ee
Defining the new variable $p = |\vec{k}| r$, 
(\ref{int}) becomes an exclusive 
function of the radial coordinate  
and reads 
\vspace{-0.2cm}
\be
\hspace{-0.5cm}\Phi(r) = - \frac{G_N M}{r} 
\frac{2}{\pi} \int_0^{+\infty} \!\!\! d p \, J_0(p) \, 
V(p^2/r^2 \Lambda^2), 
\label{int2}
\ee
which can be evaluated for the two classes of form factors here examined.

\begin{figure}[ht]
\begin{center}
\vspace{1cm}
\hspace{-0.4cm}
\includegraphics[width=4.2cm,angle=0]{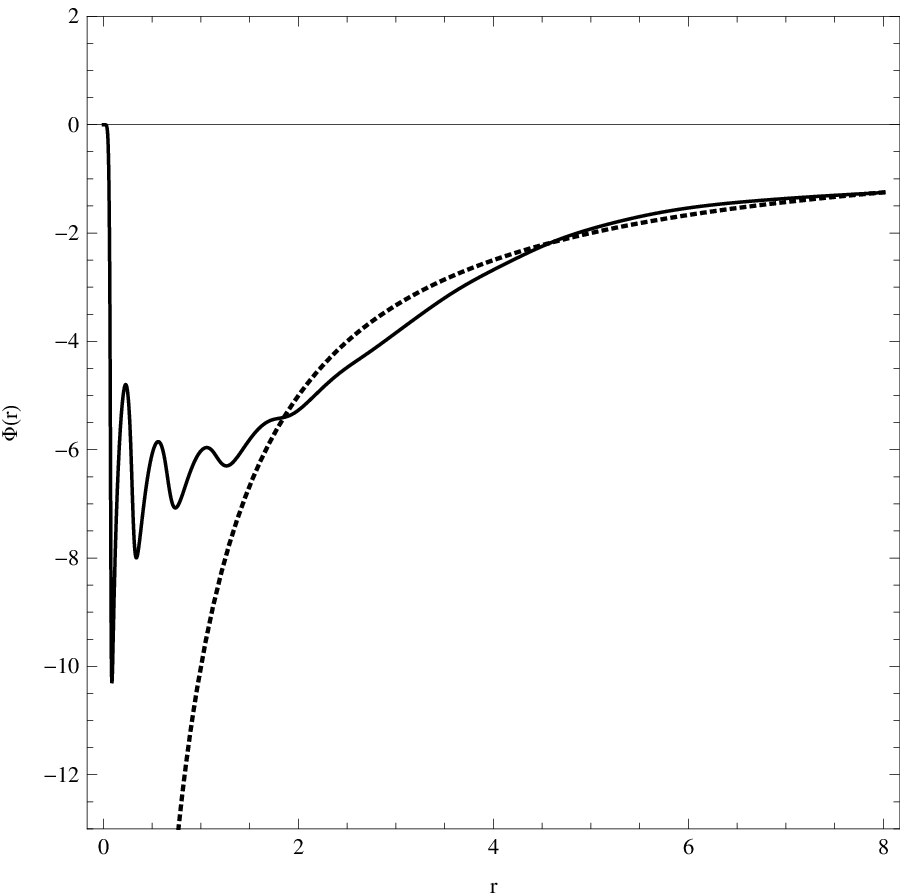}
\includegraphics[width=4.2cm,angle=0]{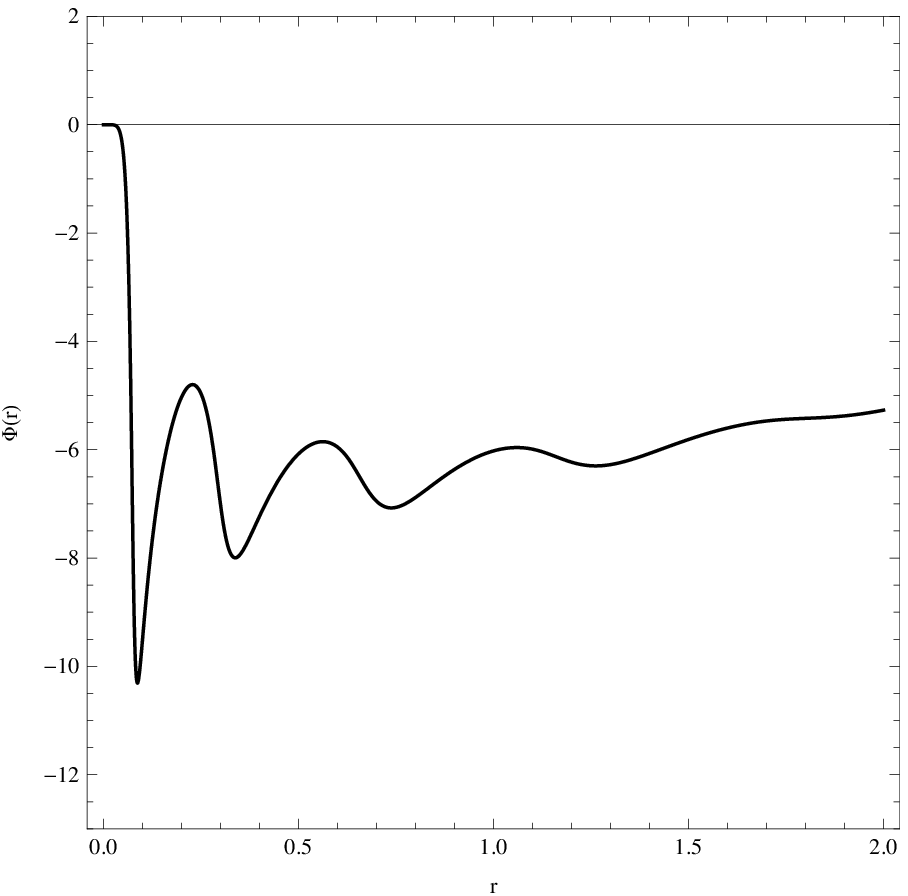}
\caption{\label{pote} Plot of the gravitational potential for $\gamma =3$ and $M=10$ ($\Lambda = G_N =1$). 
In the first plot the radial coordinate is in the range $r \in [0, 8]$ (in Planck units) and in the second one 
it is in the range $r \in [0,2]$.}
\end{center}
\end{figure}

We first evaluate the integral (\ref{int2}) for the class of theories (\ref{H}) by which $V(z) = \exp( - H(z))$.  
For small values of the radial coordinate ``$r$" (large values of ``$p$") we get 
$\Phi \approx - 2 G_N M \, ({\rm const.}) \, \Lambda^{2 \gamma+2} \, r^{2 \gamma +1}$
(where ${\rm const.} = 3 \times 10^7 \, \pi$ for $\Lambda =1$ and $G_N =1$), which is regular for 
$r\rightarrow 0$.
A plot of the exact potential for $\gamma =3$ and $M =10$ is given in Fig.\ref{pote}.

For the second class of theories with propagator given in (\ref{expp}) and $n=1$ the result 
of the integral (\ref{int}) is particularly simple,
\be
\Phi(r) = - \frac{ G_N M}{r} {\rm Er}\left(\frac{r \, \Lambda}{2} \right).
\ee
which is regular in $r=0$ and $\Phi(0) = - G_N M \Lambda/\sqrt{\pi}$. For $n>1$ the potential is again regular in $r=0$ and $\Phi(0) \propto - G_N M \Lambda$. 
We can infer that the gravitational potential is regular in the modified renormalizable theories here 
proposed.




\end{document}